\documentclass[12pt]{article}
\def\slash#1{\setbox0=\hbox{$#1$}#1\hskip-\wd0\hbox to\wd0{\hss\sl/\/\hss}}
\usepackage{epsf, cite, amsmath, amssymb}
\usepackage{epsfig}
\makeatletter
\renewcommand\section{\@startsection {section}{1}{\z@}%
                                   {-3.5ex \@plus -1ex \@minus -.2ex}
                                   {2.3ex \@plus.2ex}%
                                   {\normalfont\large\bfseries}}
\renewcommand\subsection{\@startsection{subsection}{2}{\z@}%
                                     {-3.25ex\@plus -1ex \@minus -.2ex}%
                                     {1.5ex \@plus .2ex}%
                                     {\normalfont\bfseries}}
\makeatother

\let\non\nonumber

\newcommand{\bea}{\begin{eqnarray}}
\newcommand{\eea}{\end{eqnarray}}
\newcommand{\be}{\begin{equation}}
\newcommand{\ee}{\end{equation}}

\newcommand{\hlf}{\frac{1}{2}}

\newcommand{\Z}{{\mathbb Z}}

\newcommand{\M}{{\cal M}}

\newcommand{\al}{\alpha}

\newcommand{\B}{{\mathcal{B}}}

\newcommand{\LRR}{\Longrightarrow}
\newcommand{\w}{\wedge}
\newcommand{\mo}{\omega}

\newcommand{\C}[1]{$(\ref{#1})$}

\newcommand{\CP}{{\mathbb P}}
\newcommand{\RA}{\Rightarrow}
\newcommand{\ra}{\rightarrow}

\begin{document}
\begin{titlepage}

\begin{center}

May 1, 2004
\hfill                  hep-th/0405011

\hfill EFI-04-15

\vskip 2 cm
{\Large \bf A Barren Landscape?}\\
\vskip 1.25 cm { Daniel Robbins\footnote{email: robbins@theory.uchicago.edu} 
 and  Savdeep
  Sethi\footnote{email: sethi@theory.uchicago.edu}
}\\
{\vskip 0.75cm
Enrico Fermi Institute, University of Chicago,
Chicago, IL 60637, USA\\
}

\end{center}

\vskip 2 cm

\begin{abstract}
\baselineskip=18pt

We consider the generation of a non-perturbative superpotential in
F-theory compactifications with flux. We derive a necessary condition
for the generation of such a superpotential in F-theory. 

{}For models with a single
volume modulus, we
show that the volume modulus is never stabilized by either abelian
instantons or gaugino condensation. 
We then comment on how our analysis extends to a larger
class of compactifications. From our results, it appears that among large
volume string compactifications, 
metastable de Sitter vacua (should any exist) are non-generic.

\end{abstract}

\end{titlepage}

\pagestyle{plain}
\baselineskip=18pt

\section{Introduction}

The space of four-dimensional N=1 string compactifications is a
mysterious, rich, but still largely unexplored region. 
These compactifications constitute
our current best hope of connecting string theory with observed
phenomenology. However, there are a number of basic issues that
need to be overcome. The first issue is the construction of
string vacua with, ideally, no massless scalar fields. Unfortunately,
most string vacua have many massless scalar fields, or moduli, that parametrize
the ways in which the compactification space can be deformed while
preserving supersymmetry. The second issue is breaking supersymmetry
without generating an enormous cosmological constant. It is worth
stressing that these issues are not special to string theory, but are
present in any model of low-energy phenomenology that involves
compactifying a higher dimensional theory. 

It is very possible that these problems are simply telling us that
we are still missing critical ingredients in trying to connect string
theory with observation; perhaps, the physics of the initial
cosmological singularity. However, there is another alternative first studied
in field theory~\cite{Abbott:1985qf, Brown:1988kg}. Namely, that there
exist many metastable de Sitter vacua with a distribution of
cosmological constants. By membrane nucleation, we tunnel between
vacua until we arrive at a metastable vacuum that describes our
universe. In string theory, this approach has been studied
in~\cite{Feng:2000if, Bousso:2000xa}. The typical assumption in this
kind of analysis is that metastable string solutions
exist in regions beyond our current computational control;
for example, in small volume compactifications where supergravity
cannot be trusted.

A large class of N=1
compactifications can be described in the framework of
F-theory~\cite{Vafa:1996xn}. These purely geometric
compactifications can be studied in a large volume limit, but they
tend to suffer from large numbers of moduli. A more
interesting class of vacua are found by considering F-theory compactifications
with flux described in~\cite{Dasgupta:1999ss}\ based on the M-theory
compactifications of~\cite{Becker:1996gj}. As shown
in~\cite{Dasgupta:1999ss, Gukov:1999ya}, these warped compactifications
typically have far fewer moduli
than conventional geometric compactifications. The underlying reason
for the existence of these warped, reduced
moduli compactifications is, however, a purely stringy one: the
existence of a tadpole
for D3-brane charge in F-theory~\cite{Sethi:1996es}. 

What concerns us in this letter is the structure of
non-perturbative contributions to the space-time superpotential. These
corrections have been studied in~\cite{Witten:1996bn}\ where a
criterion for non-vanishing instanton contributions was
derived. Our goal is to extend this analysis to warped compactifications with
flux. We are motivated, in part, by an interesting but yet unrealized
proposal to fix all the moduli of a flux
compactification in a regime where supergravity is valid~\cite{Kachru:2003aw}. 
Recently, there has been a summary of potential problems in scenarios
of this kind~\cite{Dine:2004fw, Banks:2003es}. However, these issues are 
secondary to the more basic question of whether any metastable vacua
actually exist.  

The proposal of~\cite{Kachru:2003aw}\ involves a set of reasonable
ingredients. Fluxes can, in principle, freeze all the geometric moduli
except for the volume modulus~\cite{Dasgupta:1999ss, mytalk}. Freezing
the volume modulus, however, requires a non-perturbative stabilization
mechanism beyond anything visible in supergravity. 
Usually, this mechanism involves abelian instantons, or
gaugino condensation. From the perspective of low-energy supergravity,
we might expect this to be a common occurence in the space of N=1
compactifications. However, string theory is not supergravity. Generic
superpotentials are not necessarily realized in string
compactifications. There are many cases that illustrate this point;
for example, 
heterotic
string vacua  are generically believed to be
destabilized by world-sheet instantons~\cite{Dine:1986zy}. However,
recently it has been shown that a large class of these vacua are
actually stable~\cite{Basu:2003bq, Beasley:2003fx, Silverstein:1995re}. 

In this letter, we
will present a kind of F-theory analogue of this result. We will
derive a condition for the generation of a non-perturbative
superpotential in F-theory (this includes certain type IIB
orientifolds~\cite{Sen:1996vd}). We then
consider the simplest class of M-theory compactifications with only
two
volume moduli, but which admit an F-theory lift. This class of models has been the subject of recent
investigations; see, for example,~\cite{Iizuka:2004ct, Giryavets:2004zr}. 
In this case, we show that neither abelian
instantons nor gaugino condensation stabilize the volume. This kind of
analysis can also be performed in a much wider class of F-theory
models with more than one volume modulus, and we present an
example. Indeed, it should be possible
to analyze most F-theory models that admit non-abelian gauge
symmetry. However, that is a subject to be explored
elsewhere~\cite{inprogress}.  

\vskip 0.2in
\noindent
{\bf Note Added:} After submitting this paper, we became aware of some
interesting work,~\cite{grassi}\ and~\cite{Denef:2004dm},  with partial 
overlap.

\section{Instantons and the Volume Modulus}

\subsection{Compactifications with flux}

We want to describe $4$-dimensional string vacua with N=1
supersymmetry. We take space-time to be flat Minkowski space. 
The class of vacua that we will consider are termed
F-theory compactifications~\cite{Vafa:1996xn}. Although string theory
is $10$-dimensional,
strangely enough, F-theory employs an 8-dimensional compactification space. 

The way this comes about is as follows: an F-theory compactification
is simply type IIB string theory compactified on a 6-dimensional
space, $\B$, with positive Ricci
curvature. To compensate the non-vanishing curvature, the type IIB
string coupling, $\tau$, must vary over $\B$. This variation  is nicely
captured in the geometry of an
8-dimensional Calabi-Yau space, $\M$, with a torus fibration. 
The base of this fibration is $\B$, while $\tau$
of the torus determines the type IIB string coupling. This structure
can be generalized to include integral NS-NS and RR 3-form
fluxes. These fluxes, denoted $H_3$ and $F_3$ respectively, 
combine into the usual IIB complex flux, $G_3$, which to preserve supersymmetry
satisfies an imaginary anti-self-duality condition~\cite{Dasgupta:1999ss}
\be
\ast G_3 = -i G_3. 
\ee  
As noted in the introduction, these compactifications have a stringy deficit
of D3-brane charge. To satisfy this Gauss constraint, we must add a
combination of D3-branes
and fluxes chosen to satisfy  tadpole
cancellation~\cite{Sethi:1996es, Dasgupta:1999ss},  
\be
{1\over 2}\int H_3 \wedge F_3 + N_{D3} = {\chi(\M) \over 24},  
\ee
where $N_{D3}$ is the number of D3-branes, while $\chi(\M)$ is the Euler
character of the 8-dimensional space, $\M$. Since adding D3-branes
introduces additional moduli, we will restrict to models with only
flux. For special choices of
$\M$, F-theory compactifications can be described as type IIB
orientifolds~\cite{Sen:1996vd}. The same is true when fluxes are
present~\cite{Dasgupta:1999ss}\ although it is worth noting that the
resulting IIB orientifolds are not necessarily perturbative string
compactifications.  

It is often useful to think about these $4$-dimensional
compactifications as limits of M-theory compactified to $3$ dimensions
on $\M$. To return to $4$ dimensions, we take the area of the torus
fiber to zero. In this limit, M-theory goes over to type IIB string
theory. Since all F-theory compactifications arise as limits of
M-theory, we can freely use either picture. This will be
useful later. The last point we should stress is that these flux compactifications
are {\it always} non-generic. Finding supersymmetric flux vacua requires
fine-tuning both complex structure and K\"ahler structure moduli
(see~\cite{Dasgupta:1999ss}\ for details). This is the
reason that many scalars are frozen. Indeed, in some orbifold examples
of~\cite{Dasgupta:1999ss}, and some $K3\times K3$ examples
of~\cite{Dasgupta:1999ss, Tripathy:2002qw}, all the complex structure
moduli are fixed. However, the overall volume modulus is never fixed
this way.   

\subsection{Abelian instantons}

Now let us turn to the generation of a non-perturbative
superpotential. It is simplest to begin with smooth spaces $\M$ that only
give rise to abelian symmetry. We will explain later that this is actually
the only case that ever needs to be studied. 

Since all F-theory compactifications are limits of M-theory
compactifications, we begin in the more general M-theory setting. In
this setting, the background flux, denoted $G_4$, is a (half) integral self-dual
primitive $4$-form. The integrality of $G_4$ is correlated 
with $\chi/24$~\cite{Witten:1997md}.  
{}From the M-theory perspective, instantons 
are constructed by wrapping Euclidean M2 and M5-branes on cycles of $\M$. To
stabilize volume moduli, we are interested in M5-brane instantons which,
in the absence of $G_4$ flux, wrap complex divisors, $D$, of $\M$. To generate
a term in the superpotential, an instanton must produce exactly the right
number of fermion zero-modes. In~\cite{Witten:1996bn}, Witten
derived a necessary condition on $D$ for this to be the
case, 
\be\label{arithmetic}
\chi_D = 1,
\ee
where $\chi_D$ is the arithmetic genus of the divisor. This condition
is necessary but not sufficient. It can be satisfied by divisors which
have moduli. Whether those divisors contribute is hard to
determine. The only divisors that contribute for sure are rigid (in
the absence of D3-branes~\cite{Ganor:1998ai}). We
will actually rule out any divisor satisfying~\C{arithmetic}. 

To account for the background flux, we will take the following
approach. At the level of the supergravity solution, we are free to
ignore flux quantization, and 
tune the flux to zero. The metric on $\M$, which is conformally
Calabi-Yau~\cite{Becker:1996gj}, becomes simply Calabi-Yau and we
arrive at a standard M-theory compactification. The warp
factor becomes constant.   Under this smooth deformation, a 
BPS instanton should remain BPS, and so must satisfy~\C{arithmetic}.  

This does not mean that the $G_4$ flux is irrelevant! In fact, it is harder
to generate a superpotential in flux compactifications because not
all divisors contribute. The M5-brane world-volume contains a coupling
between $G_4$ and the anti-self-dual world-volume 2-form, $b_2$. This
coupling is proportional to
\be
\int_{D} b_2 \wedge G_4. 
\ee
Wrapping a Euclidean M5-brane on a divisor through which $G_4$ flux
threads generates a source for $b_2$. For a compact divisor, these
sources must sum to zero so the total flux through $D$ must
vanish.

\subsection{Reformulating the arithmetic genus condition}

So now we are in the situation of trying to find divisors, $D$, in
$\M$ with $\chi_D=1$. Our task is aided by the integral expression for
the arithmetic genus, 
\be
\chi_D = \int_D {\rm Td}(D),
\ee
where ${\rm Td} (D)$ is the Todd class of $D$. In terms of Chern classes, the
Todd class is given by,
\be
{\rm Td}(D) = 1+\hlf c_1+\frac{1}{12}\left[c_2+c_1^2\right]+\frac{1}{24}c_2c_1.
\ee
We denote the $(1,1)$ cohomology class Poincar\'e dual to
$D$ by $[D]$. This form acts like a delta function restricting us to
$D$ so 
$$\int_D\eta=\int_\M \eta\w[D]$$
{}for all forms $\eta$ on $\M$.
Now by the adjunction formula, the total Chern 
class of $D$ is given in terms of the Chern classes of $\M$ by
\be
c(D) = \frac{c(\M)}{1+[D]} = 1-[D]+\left(c_2(\M)+[D]^2\right)+\left(c_3(\M)-
c_2(\M)[D]-[D]^3\right),
\ee
where all forms are understood to be pulled back to $D$. 
In particular, all of the Chern classes of $D$ are pull-backs of forms
on $\M$.  Then\footnote{Alternatively, this follows directly from the Hirzebruch-Riemann-Roch formula for
this case,
\be
\chi_D = \int_\M\left(1-e^{-[D]}\right){\rm Td}(\M).
\ee
}
\be
\label{chiD}
\chi_D = \frac{1}{24}\int_D c_2(D)c_1(D) = -\frac{1}{24}\int_\M\left(c_2(\M)[D]^2
+[D]^4\right), 
\ee
and the arithmetic genus condition becomes
\be
\int_\M\left(c_2(\M)[D]^2+[D]^4\right) = -24.
\ee

\subsection{Divisors dual to the K\"ahler cone of $\M$}

An interesting case to consider is the case when $[D]$ lies inside the K\"ahler
cone of $\M$, i.e., when $[D]$ would be an acceptable choice for the
K\"ahler form of $\M$.  In this case, we immediately have that
\be
\int_\M [D]^4 = 4!\times{\mathrm{Volume}}(\M) \ge 0.
\ee
In order for such a divisor to have $\chi_D>0$ it follows from~\C{chiD}\ that 
\be
\int_\M c_2(\M)[D]^2 < 0.
\ee
However, this is impossible by the following inequality on a Calabi-Yau
$n$-fold $X$ with K\"ahler class $[\mo]$~\cite{MR88f:53087},\footnote{This is a special
  case of a more general inequality which
holds for Einstein-K\"ahler manifolds,
\be
\int_X \left(c_1(X)^2-\frac{2(n+1)}{n}c_2(X)\right)[\mo]^{n-2}\le 0.
\ee} 
\be
\int_Xc_2(X)[\mo]^{n-2} \ge 0.
\ee
So if $D$ is dual to a K\"ahler class then instantons wrapping $D$ cannot
contribute to the superpotential.   

As an immediate application of this result, consider an $\M$ that has only one
K\"ahler modulus which parametrizes the overall volume of $\M$. 
In this case {\it{every}} effective divisor must lie in the K\"ahler cone
since the K\"ahler class necessarily spans $H^2(\M)$ (an effective divisor in
this case will be a positive multiple of the K\"ahler class which is
also integral, but we do not need this here).  So in this case, the
volume cannot be stabilized by abelian instantons. Models with one
K\"ahler modulus cannot be elliptic fibrations, so these models only
exist as $3$-dimensional M-theory compactifications. 
To rule out F-theory models with one K\"ahler
modulus, we need to do more work. 

\subsection{An F-theory condition on instantons}
\label{ftheorycondition}
{}To study compactifications with a type IIB description, we want $\M$ to
be an elliptically-fibered Calabi-Yau $4$-fold with base $B$. We can
present $\M$ in  Weierstrass form as follows: let $W$ be a $\CP^2$
bundle over $B$ with homogeneous
coordinates $(x,y,z)$ which are sections of ${\mathcal O}(1)\otimes K^{-2}$, 
${\mathcal O}(1)\otimes K^{-3}$, and ${\mathcal O}(1)$,  respectively. The line-bundle
${\mathcal O}(1)$ is the degree one
bundle over the $\CP^2$ fibre and $K$ is the canonical bundle of $B$. Then
$\M$ is given by 
\be
0 = s = -zy^2+x^3+axz^2+bz^3,
\ee
where $a$ and $b$ are sections of $K^{-4}$ and $K^{-6}$ and $s$ is a section 
of ${\mathcal O}(3)\otimes K^{-6}$.  Now the cohomology ring of $W$ is given by adding 
the element $\al=c_1({\mathcal O}(1))$ to the cohomology ring of $B$ along with the 
relation,
\be
0 = \al(\al+2c_1)(\al+3c_1),
\ee
where $c_1=c_1(B)$.  This relation follows because $x$, $y$, and $z$ are not
allowed to have any common zeroes.  In the cohomology ring of $\M$, there is a
further simplification. Since $\M$ is given by $s=0$, we always multiply by
$c_1({\mathcal O}(3)\otimes K^{-6})=3(\al+2c_1)$ before integrating, so the relation
above can be simplified to
\be
0 = \al(\al+3c_1) \quad \RA \quad \al^2=-3\al c_1
\ee
on $\M$.

Write the total Chern class of $B$ as $c(B) = 1+c_1+c_2+c_3$.  Now the total 
Chern class of $\M$ is calculated by adjunction
\bea
c(\M) &=& \left(1+c_1+c_2+c_3\right)
\frac{(1+\al+2c_1)(1+\al+3c_1)(1+\al)}{1+3\al+6c_1}\non\\
&=& 1+(c_2+11c_1^2+4\al c_1)+
(c_3-c_1c_2-60c_1^3-20\al c_1^2)+4\al(c_1c_2+30c_1^3).
\eea

In M-theory, superpotentials are generated by M5-branes wrapping
divisors with arithmetic genus one,
$\chi(D)=1$ given by~\C{chiD}.   
In terms of the $(1,1)$ cohomology class of the divisor, $[D]$, and the 
expression derived above for the Chern class of $\M$, we can rewrite
the arithmetic genus constraint as
\be
[D]^4+\left(c_2+11c_1^2+4\al c_1\right)[D]^2 = -24.
\ee

Finally, we would like to take the F-theory limit and relate this to a
compactification of type IIB string theory on $B$.  Let $\pi:\M\ra B$ be the
projection onto the base.  If $\pi(D)=B$ then the
contribution to the superpotential vanishes on taking the F-theory limit, so
we need only restrict ourselves to the case that $D=\pi^{-1}(C)$, where $C$ is
a divisor in the base $B$.  In particular,  this implies that 
\be [D]_\M^4=\pi^{-1}([C]_B^4)=0
\ee 
and also that
\be (c_2[D]^2)_\M=\pi^{-1}\left( (c_2 [C]^2)_B \right)=0, \qquad
\left( c_1^2[D]^2 \right)_\M=0. 
\ee  
So for this situation,  the constraint reduces greatly to
read
\be \label{reduced}
4\al c_1[D]^2 = -24.
\ee
Finally, we can do the fiber integration by multiplying~\C{reduced}\ by $3(\al+2c_1)$
and picking out the $\al^2$ term, leaving only a condition on the base
\be
\label{D3cond}
c_1[C]^2 = -2.
\ee

One immediate application of this formula is to show that in cases with only
one K\"ahler modulus, there can be no superpotential generated by smooth
instantons.  Indeed, if $h^{1,1}(B)=1$, then the K\"ahler form $J$ generates
all of $H^{1,1}(B)$.  In particular, we know that $c_1(B)$ is a positive 
$(1,1)$-form, since $c_1(B)=c_1(K^{-1}(B))$, and (e.g.) $K^{-4}$ had at least
one nonzero section, so $c_1=aJ$ for some $a>0$.  We also know that $[C]=bJ$ for
some real (in fact positive, but we do not actually require that) $b$.  Then
\be
c_1[C]^2 = ab^2\int_BJ^3 = 3! \times ab^2\times Vol(B)\ge 0.
\ee
In particular, no divisors $C$ can satisfy the condition
(\ref{D3cond}). Therefore, the volume is not stabilized in F-theory
models with one K\"ahler modulus.\footnote{We should note that the condition
$h^{1,1}(B)=1$ does not imply that $h^{(1,1)}(\M)=2$. There can be
more than two K\"ahler moduli on $\M$ which arise from reducible singularities
in the elliptic fibration. None of these models can be stabilized by
our argument. We wish to thank P.~Aspinwall and the Duke CGTP group
for this observation.}

\subsection{Gaugino Condensation}

Another natural stabilization mechanism contemplated
in~\cite{Kachru:2003aw}\ is a superpotential generated by gaugino 
condensation. Non-abelian gauge symmetry arises in F-theory via
coincident $7$-branes which wrap a divisor $S$ of $B$. By an
$SL(2,\Z)$ transformation, we can choose these $7$-branes to be
D7-branes. 
In M-theory,
this gauge symmetry comes about via an $ADE$
singularity fibered over $S$. We want models
with pure N=1 Yang-Mills so we need to freeze the scalars on the
D7-branes. This can be achieved in two ways: the first is by choosing
an $S$ so that~\cite{Katz:1997th} 
\be
h^{1,0}(S) = h^{2,0}(S)=0. 
\ee    
This condition ensures that there are no moduli from either
Wilson lines on $S$, or from the twisted scalars on the D7-branes. In
this situation, the D7-branes simply cannot move, and we get pure
gauge symmetry. There is one known F-theory compactification to $6$
dimensions that has a pure gauge factor of this kind~\cite{Bershadsky:1996nh}. 

The other possibility is to consider coincident motile D7-branes. This notion
is really only precise at points where all the branes
are mutually local D7-branes like the orientifold
point~\cite{Sen:1996vd}. We can try to freeze the moduli of
motile D7-branes using flux. In this case, there are moduli from
either $h^{1,0}(S)$ or $ h^{2,0}(S)$. The former correspond to
moduli from 
$h^{2,1}(\M)$ of the four-fold $\M$, while the latter are complex
structure deformations 
generated by $h^{3,1}(\M)$. Both classes of moduli can be frozen by
flux~\cite{Dasgupta:1999ss}. However, whether there is any unbroken
gauge symmetry depends on the kind of background $G_4$. Some choices
of $G_4$ lift to instantons embedded in the gauge group on the
D7-branes wrapped on $S$. These instantons can either partially or
completely break the gauge
symmetry. In addition, there are
moduli associated with the embedding of these instantons.   

{}Fortunately, these caveats will not affect our
conclusions.\footnote{We wish to thank Sheldon Katz for explaining the
  following argument to us.} Geometrically, we have the following projections, $\M \ra
B \ra S$. We assume that $h^{1,1}(B)=1$, and that the fibers of
$\M$ over $S$ are of $ADE$ type so that we have enhanced gauge symmetry. If
$J$ is the
K\"ahler form of $B$ then $\int_B J^3>0$. However, by construction,
$J$ is necessarily the pull back of some $(1,1)$ class $H$ on $S$. Since
$H^3=0$ on the complex surface $S$, this implies that $J^3=0$ which is
a contradiction. Since there is no space $B$ with only one K\"ahler
modulus that can give gaugino
condensation, the volume cannot be
stabilized in these models.

Actually, the entire discussion about gaugino condensation is
superfluous. As mentioned earlier, we are always free to compute the
superpotential in M-theory. Reducing an N=1
Yang-Mills vector multiplet to $3$ dimensions always results in one
adjoint-valued scalar from the Wilson line in the
direction of reduction. In $3$ dimensions, we therefore always have a
Coulomb branch, and we are free to break the gauge symmetry down to
abelian. This corresponds to resolving the $ADE$ fibers over $S$ by a
resolution parameter that depends on the area of the elliptic
fiber. The situation then reduces to the case studied in
section~\ref{ftheorycondition}, and we arrive at the same conclusion.

\subsection{More general models}

{}Finally, we will outline how our analysis extends to more general
models with many K\"ahler moduli. As a specific example of another case, consider an
elliptically-fibered Calabi-Yau $4$-fold,
$\M$, over a base $B=\CP^2\times\CP^1$.  The cohomology ring of 
the base is generated by the K\"ahler classes of the two factors, which we 
denote by $J$ and $\beta$ for $\CP^2$ and $\CP^1$,
respectively. These classes obey the relations,
$J^3=\beta^2=0$.  We can choose $J$ and $\beta$ to generate the 
integral cohomology of $B$ so that 
\be \int_{\CP^2}J^2=\int_{\CP^1}\beta=1. \ee  
The class of an effective divisor can then be expressed in terms of
$J$ and $\beta$, 
\be
[C] = nJ+m\beta,
\ee
where $n$ and $m$ are non-negative integers.

To find the (possibly) contributing divisors, we use the
criterion~\C{D3cond}\ which means we only have to compute
\be
c_1(B)[C]^2 = \int_B\left(3J+2\beta\right)\left(nJ+m\beta\right)^2 = 2n^2+6nm
\ge 0.
\ee
So we see immediately that there are no contributing divisors in this case.

Actually, it is a simple matter to extend this analysis to the case of any
product $B=S\times\CP^1$ (this is, for instance, the case considered in~\cite{Donagi:1996yf},
where the superpotential is calculated for a specific $S$).  
Here $S$ must itself be a $c_1$ positive surface
with arithmetic genus one so that $B$ also has these properties.  Let 
$\beta$ again denote the integral class of $\CP^1$.  Then we have
$c_1(B)=c_1(S)+2\beta$ and we can write any effective divisor as 
$[C]=[E]+k\beta$ where $[E]$ is a divisor on $S$ and $k$ is a non-negative 
integer.  We again compute
\bea
c_1(B)[C]^2 &=& \int_B\left(c_1(S)+2\beta\right)\left([E]+k\beta\right)^2\non\\
&=& \int_B\left(2[E]^2+2kc_1(S)[E]\right)\beta = 2\int_S[E]\left([E]+kc_1(S)
\right).
\eea

Since $c_1(S)$ is a positive integral class on $S$, $[E]$ and $[E]+kc_1(S)$ 
both represent effective divisors on $S$.  If $[E]$ is a multiple of $c_1(S)$ 
(this was the case in the $S=\CP^2$ example above) then this is proportional 
to $\int_Sc_1(S)^2\ge 0$.  If $[E]$ is not proportional to $c_1$ and $k>0$ 
then $[E]$ and $[E]+kc_1(S)$ really represent two different curves in $S$ and 
hence intersect non-negatively.  So the only remaining case is that $k=0$ and 
$E$ has negative self-intersection.  For a smooth abelian instanton 
contribution, we see that the condition is in fact $\int_S[E]^2=-1$.

Moreover, if $[E]$ is irreducible, then by adjunction again we see that
\be
c(E) = \frac{c(S)}{1+[E]} = 1+\left(c_1(S)-[E]\right),
\ee
and hence
\bea
\chi(E) &=& \int_Ec_1(E) = \int_S\left(c_1(S)-[E]\right)[E]\non\\
\LRR \int_S[E]^2 &=& \int_Sc_1(S)[E]-\chi(E) = \int_Sc_1(S)[E]-2+2g,
\eea
where $g$ is the genus of the curve $E$.  Since the integrand on the right
hand side above is again non-negative, as is $g$, we have a negative total
result only for $g=0$ and $\int_Sc_1(S)[E]<2$.  To get a contributing divisor 
we need specifically $\int_Sc_1(S)[E]=1$.

It is not difficult to construct such examples. For instance if $S$ is a del
Pezzo surface or the Hirzebruch surface $\mathbb F_1$.  However, since
all contributing divisors wrap the $\CP^1$, their volumes are invariant if
we scale the size of $\CP^1$ by a factor $t$ while also scaling the surface
$S$ by a factor $t^{-1}$.  But the non-perturbative superpotential depends only
on the volumes of the contributing divisors, and hence is independent of the
K\"ahler modulus associated with this deformation.  So we again conclude that there
is at least one K\"ahler modulus unfixed by these non-perturbative effects.

It is possible to perform a similar analysis in many even more general classes
of examples~\cite{inprogress}.  This will extend prior work
on superpotentials in M and F-theory~\cite{Witten:1996bn,
  Donagi:1996yf,  Mayr:1997sh, Klemm:1998ts,
  Braun:2000hh}\ to broad classes of non-singular
spaces. However, what seems clear is that complete moduli
stabilization in string theory is hard to achieve!

\section*{Acknowledgements}

It is our pleasure to thank Jeff Harvey, Ken Intriligator, Anatoly
Libgober, Joe
Polchinksi, Paul Seidel, and
particularly, Sheldon Katz for helpful discussions. S.~S. would like
to thank the Aspen Center for Physics for hospitality. 
The work of D.~R. is supported in part by a 
Julie Payette--NSERC PGS B
Research Scholarship, and by a Fermi-McCormick Fellowship.
The work of S.~S. is supported in part by NSF CAREER Grant
No. PHY-0094328, and by the Alfred P. Sloan
Foundation.

\newpage




\providecommand{\href}[2]{#2}\begingroup\raggedright\endgroup
\end{document}